# Superconducting and low temperature RF Coils for Ultra-Low-Field MRI: A Study on SNR Performance

Aditya A Bhosale[1], Komlan Payne[1], Xiaoliang Zhang[1,2*]

[1]Department of Biomedical Engineering, State University of New York at Buffalo, Buffalo, NY, United States

[2]Department of Electrical Engineering, State University of New York at Buffalo, Buffalo, NY, United States

[*]Corresponding Author:

Xiaoliang Zhang, Ph.D.
Deaprtment of Biomedical Engineering
SUNY Buffalo
215E Bonner Hall
Buffalo, NY 14260
USA
Email: xzhang89@buffalo.edu

# Abstract

Ultra-low field MR imaging systems are becoming increasingly popular due to their affordability, portability, compact size, utility in monitoring brain tumors during treatment, structural imaging for early stroke detection, and guiding interventional procedures. However, these systems often face challenges such as low contrast, low spatial resolution, and poor image quality, which can impact diagnostic precision and treatment planning. While artificial intelligence algorithms and denoising techniques have been proposed to improve the signal-to-noise ratio (SNR), these methods often come with trade-offs, including increased computational complexity, potential loss of image fidelity, and challenges in result validation.

High-conductivity materials, particularly high-temperature superconductors (HTS), have shown significant promise in the RF hardware design for magnetic resonance imaging (MRI). These materials, with their nearly infinite conductivity, have successfully improved SNR at low fields of one-tenth or a few tenths of a Tesla up to high fields of 3 Tesla. However, their effectiveness at ultra-low fields, such as 70mT, a new regime for human MRI with numerous promising applications, remains largely uninvestigated.

This study incorporates electromagnetic simulations to assess the performance of multi-turn solenoid coils for ultra-low field MR imaging with various conductor materials (superconducting material, low-temperature copper, and room-temperature copper) across different human samples (elbow, knee, and brain). At 70 mT, superconducting materials performed significantly better than both room-temperature and low-temperature copper. The high Q-factor of the superconducting material indicates lower energy loss, which is useful for MR imaging. Furthermore, B1+ field efficiency increased significantly with superconducting materials, indicating superior performance. SNR evaluations revealed that materials with higher conductivity significantly improve SNR, which is critical for producing high-quality MR images. These results show that superconducting and low-temperature copper materials can significantly improve MR imaging quality at ultra-low fields, which has important implications for coil design and optimization.

# 1. Introduction

Ultra-low field (~70mT) MR imaging systems are gaining popularity due to their advantages, including lower costs that make them accessible to a wider population, structural imaging capabilities for onsite monitoring of brain tumor treatment, reduced shielding requirements and portability due to their light weight, and ongoing research into their potential applications in interventional procedures [1-5]. These systems are also considered safer due to their low power requirements [6-8]. Despite these benefits, ultra-low field MRI systems face significant challenges compared to high and ultrahigh field systems which have demonstrated significantly enhanced sensitivity, spatial resolution and spectral dispersion [9-21]. These challenges are evident in poor image quality resulting from a low signal-to-noise ratio (SNR), reduced spatial resolution, and limited contrast issues that are particularly problematic in clinical scenarios requiring precise anatomical details or subtle pathological features. For instance, in neuroimaging, accurate localization of small lesions or subtle abnormalities is critical, and the limitations of ultra-low field MRI can hinder diagnostic accuracy and treatment planning. Additionally, the acquisition process is prolonged due to the necessity of increased averaging to elevate the SNR. The use of imaging acceleration techniques, such as, deep learning, compressed sensing, and parallel imaging, may potentially mitigate this problem and facilitate a more rapid acquisition [22-33].

Efforts to improve SNR using artificial intelligence (AI) algorithms and denoising techniques show promise but are not fundamental improvements and are often facing challenges in validating their outcomes. Although these post-processing methods can mitigate some SNR issues, they often introduce trade-offs. These include increased computational complexity, potential loss of image fidelity, and missing information on small lesions or anatomic structures due to the limited resolution of the raw data. Moreover, these approaches might not address the underlying hardware limitations contributing to SNR degradation, underscoring the need for alternative RF hardware solutions.

Low temperature copper and superconducting materials, such as High-Temperature Superconductors (HTS), offer the exciting possibility of high or near-infinite conductivity, presenting a promising alternative to traditional materials in reducing the conductor losses of RF coils. It has been reported that low temperature RF coils and superconducting RF coils, due to their reduced conductor losses, can enhance SNR [34, 35] in MRI at field strengths ranging from one-tenth or a few tenths of a Tesla up to 3T, where the noise of MR is dominated by the noise of RF coils [36-38]. In the high and ultrahigh field regime at 4T and above [39-46], the increased frequency results in the emergence of sample noise domination behavior in human MR [47-52]. Consequently, the effectiveness of low temperature or superconducting RF coils in improving MR SNR is diminished. However, their effectiveness at ultra-low field strengths, such as 70mT, remains largely uninvestigated and has yet to be fully elucidated. This study aims to ascertain whether superconducting or low temperature RF coils can improve SNR at such low fields and, if so, to quantify the potential gains. This investigation will guide decisions on the use of superconductors or low temperature in RF hardware for ultra-low field MRI applications.

In this work, we investigate the performance of low-temperature copper and superconducting materials through the commonly used multi-turn solenoid coils in the ultralow field regime (~70mT). All the coils are tuned to 3MHz, the proton Larmor frequency at the ultra-low field of 70mT [53]. Our assessment includes SNR, $B1^+$ efficiency, and Q-factor evaluations based on different material properties. By rigorously analyzing the impact of material selection on SNR at ultra-low field strength of 70mT, we aim to provide valuable insights into optimizing RF hardware and to offer guidance on RF coil design strategies for ultra-low field MRI applications.

# 2. Methods

## 2.1 Multi-turn Solenoid Coil

The multi-turn solenoid coil was modeled and simulated using CST Microwave Studio (CST MWS) software. The coil consisted of 16 turns of 3mm wide tape as the conductor, with 12.75 mm spacing between the turns. The total length of the coil was 255 mm, and its diameter was 260 mm. The design and dimensions of the coil are illustrated in Figure 1. The coil was designed to operate at approximately 3 MHz to operate at a magnetic field strength of 70 mT for MR imaging. To tune the coil to the desired frequency, a single tuning capacitor was used. Additionally, a matching capacitor was connected parallel to the feed port to match the impedance to 50 ohms, ensuring efficient power transfer. The performance of the coil was evaluated by analyzing loaded and unloaded Q-factors, $B1^+$ efficiency maps in the imaging sample, noise estimation, SNR 3D maps, and 1D profiles in the imaging sample.

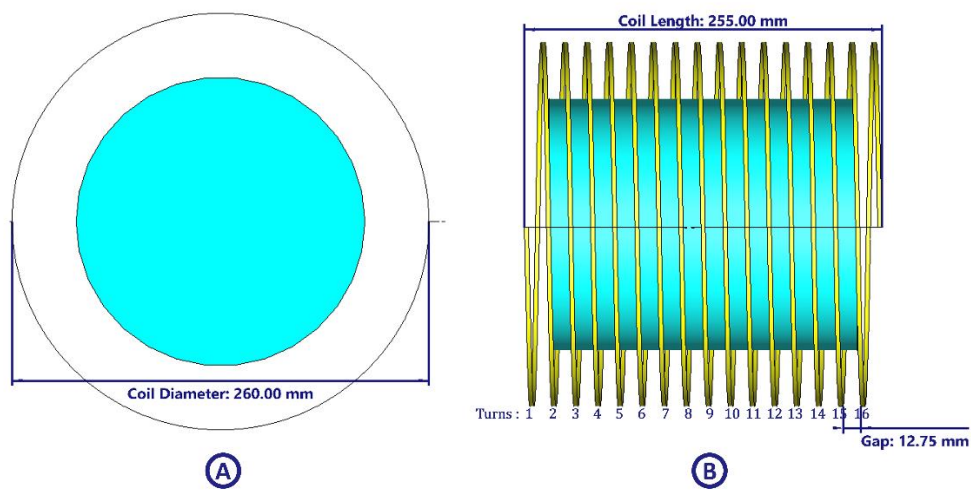

*Figure 1.* (a) Top view of the 16-turn solenoid coil showing its diameter. (b) Side view of the solenoid coil illustrating the coil length, gaps between each turn, and labeling of each turn.

## 2.2 Coil Conductivity

The coil conductor was assigned three different conductivity materials to evaluate performance under various conditions. The first material was room temperature copper, with a resistivity of $1.68\times10^{-8}$ Ω·m and a conductivity of $5.96\times10^7$ S/m. The second material was low-temperature copper, evaluated at liquid nitrogen temperature (-193.80°C). The resistivity at this temperature was $2.79\times10^{-9}$ Ω·m, resulting in a conductivity of $37\times10^7$ S/m. Finally, a material mimicking superconducting conditions was implemented with an elevated conductivity of $10,000\times10^7$ S/m.

## 2.3 Sample

Three samples were chosen to evaluate the coil's SNR performance at different conductivities. The coil remained the same throughout the study with no changes to its shape and size. Only the samples were different, and the coil was tuned and matched at 3MHz by fine-tuning the lumped components of the coil. The samples used were: Human Elbow ($\varepsilon_r$ : 565 and $\sigma$ : 0.197 S/m), Human Knee ($\varepsilon_r$ : 251 and $\sigma$ : 0.233 S/m), and Human brain ($\varepsilon_r$ : 565 and $\sigma$ : 0.197 S/m). All samples were adult male sized, resembling actual human anatomy. The 3D perspective views of the solenoid coil with labeled material properties for different human models are shown in Figure 2.

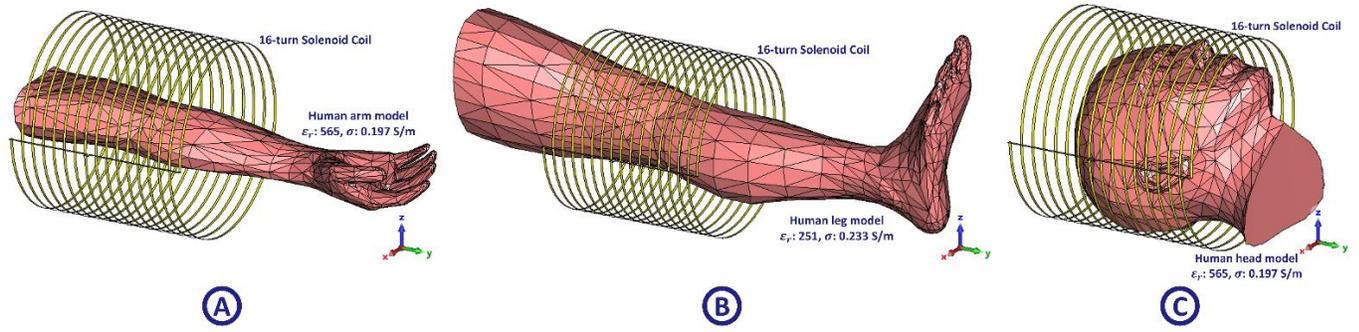

*Figure 2.* 3D perspective views of the solenoid coil with labeled material properties for different human models: (a) human arm, (b) human leg, and (c) human head.

## 2.4 Q-Factor evaluation

The Quality Factor (Q-factor) was evaluated to determine the efficiency of the multi-turn solenoid coil under different conditions. The Q-factor was calculated by dividing the resonant frequency by the -3dB frequency bandwidth. Unloaded Q-factors were assessed for the coil with different assigned conductivities, including room-temperature copper, low-temperature copper, and a material mimicking superconducting conditions. To evaluate the loaded Q-factors, the coil was tested with three different samples (human elbow, human knee, human brain) placed in the vicinity of the coil. The evaluations were conducted using numerical simulations within CST Microwave Studio (CST MWS). The results of these evaluations were essential for estimating the noise levels, which were subsequently used in the SNR calculation.

## 2.5 $B_1^+$ field evaluation

The B1+ field evaluation was performed to compare the B1+ field efficiency of the coil with different conductivities. Numerical simulations in CST Microwave Studio (CST MWS) were used to generate B1+ field maps for each conductivity material. The B1+ field was measured in units of microtesla per square root of input power ($\mu T/\sqrt{W}$).

The B1+ field maps were plotted for each sample to evaluate changes in the B1+ distribution and mean values across different conductivity materials. These maps provided a detailed view of the magnetic field strength and uniformity within the samples. The resulting B1+ field distributions were then utilized as signal inputs in the SNR evaluation to assess how different conductivity materials affect the coil's performance.

## 2.6 SNR Calcultion

The Signal-to-Noise Ratio (SNR) was calculated to evaluate the coil's performance. The calculation of noise involved several key equations [16]. The angular frequency ($\omega$) was determined using the formula: $\omega = 2\pi f$, where $f$ is the resonant frequency.

The inductive reactance ($L$) was calculated using: $L = \frac{1}{(2\pi f)^2 \times C}$, where $C$ is the capacitance.

The coil resistance ($R_c$) was computed as: $R_c = \frac{L \times \omega}{Q_{unloaded}}$, with $Q_{unloaded}$ representing the unloaded Q-factor.

The sample resistance ($R_s$) was calculated using: $R_s = \frac{L \times \omega (Q_{unloaded} - Q_{loaded})}{Q_{unloaded} \times Q_{loaded}}$, where $Q_{loaded}$ is the loaded Q-factor.

The noise was determined using the formula: $Noise = \sqrt{R_c T_c + R_s T_s}$, where $T_c$ is the coil temperature and $T_s$ is the sample temperature.

Finally, the SNR can be calculated using the following equation: $SNR \propto \frac{B1^+}{Noise}$

The SNR maps were constructed by using the B1$^+$ field maps as the signal input and the noise values calculated with the aforementioned equations. The SNR evaluations included analyzing the mean SNR values in seven 10×10 mm² regions of interest (ROIs) placed inside the samples. These ROIs were strategically positioned to provide a comprehensive assessment of the coil's performance across different areas within the samples. The mean SNR values from these regions were used to evaluate and compare the coil's efficiency and imaging capabilities with different conductivity materials.

# 3. Results

## 3.1  Q-factor evaluation

The unloaded and loaded Q-factors of the multi-turn solenoid coil were evaluated for different conductor materials and imaging samples. The unloaded Q-factors for the coil with various conductivities were as follows: room-temperature copper had an unloaded Q-factor of 218.44, low-temperature copper exhibited an unloaded Q-factor of 486.25, and the superconducting material showed an unloaded Q-factor of 5809.9 (see figure 3a).

The loaded Q-factors, measured with different samples, indicated significant variations. For the human elbow, the loaded Q-factors were 180.56 for room-temperature copper, 278.83 for low-temperature copper, and 1134.28 for the superconducting material (see figure 3b). For the human knee, the loaded Q-factors were 158.45 for room-temperature copper, 266.69 for low-temperature copper, and 652.29 for the superconducting material (see figure 3c). For the human brain, the loaded Q-factors were 138.88 for room-temperature copper, 213.99 for low-temperature copper, and 347.87 for the superconducting material (see figure 3d). These results demonstrate that the superconducting material consistently achieved the highest Q-factors, both unloaded and loaded, indicating superior performance in terms of efficiency and reduced resistive losses. The improvement in Q-factors with higher conductivity materials underscores their potential for enhancing coil performance in MR imaging applications.

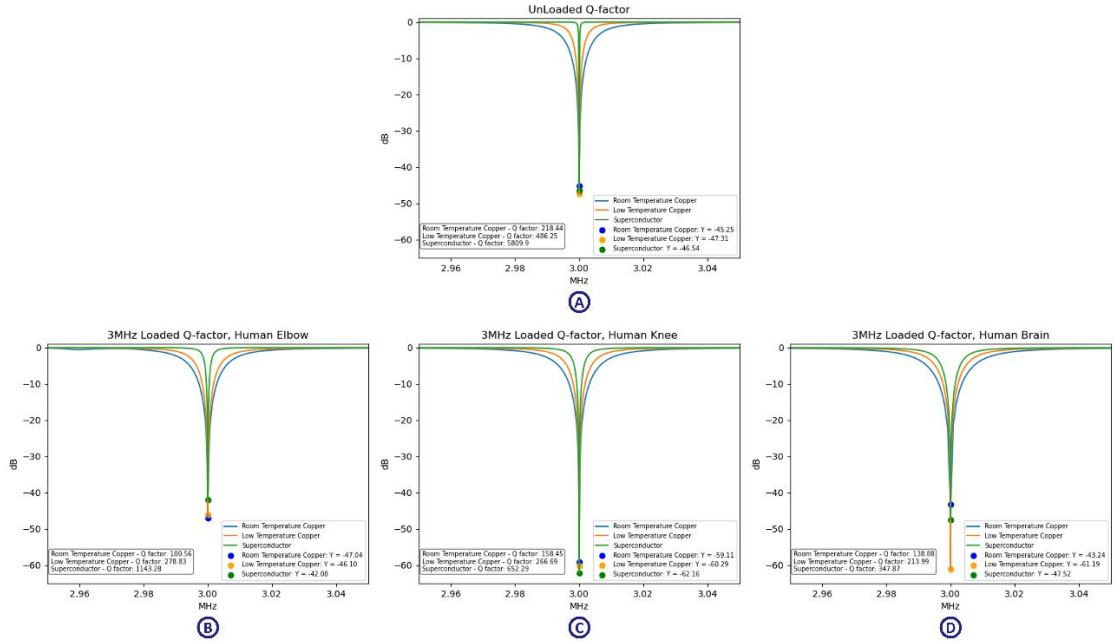

*Figure 3.* *Reflection coefficient S11 plots showing quality factors for different conductivity parameters: (a) unloaded coil, (b) coil loaded with human elbow model, (c) coil loaded with human knee model, and (d) coil loaded with human brain model.*

## 3.2   Noise Evaluation

The noise levels were evaluated based on the calculated resistance and temperature parameters, using the equations provided in the methods section. The noise values, in combination with the $B_1^+$ field maps, were used to generate SNR maps. These maps illustrated and helped compare the performance of the coil with different conductive materials under various conditions. Noise calculation parameters for various conductivity conditions are summarized in Table 1.

*Table 1.* Noise calculation parameters for various conductivity parameters across different human models.

| | | | | Human Elbow | | | | | |
|---|---|---|---|---|---|---|---|---|---|
| **Material** | **Unloaded Q-factor** | **Loaded Q-factor** | **Unloaded to Loaded Q-factor Ratio** | **Capacitance (pF)** | **Inductance (μH)** | **Coil Resistance (Ω)** | **Sample Resistance (Ω)** | **Noise** | **Sample Noise to Coil Noise Ratio** |
| **Room-temperature Copper** | 218.44 | 180.56 | 1.209791759 | 44.7 | 62.91 | 5.43 | 1.14 | 44.39 | **0.209945** |
| **Low-temperature Copper** | 486.25 | 278.83 | 1.743894129 | 46.375 | 60.99 | 2.36 | 1.75 | 26.6 | **0.741525** |
| **Superconductor** | 5809.9 | 1143.3 | 5.081693344 | 49 | 57.65 | 0.19 | 0.77 | 15.19 | **4.052632** |
| | | | | **Human Knee** | | | | | |
| **Material** | **Unloaded Q-factor** | **Loaded Q-factor** | **Unloaded to Loaded Q-factor Ratio** | **Capacitance (pF)** | **Inductance (μH)** | **Coil Resistance (Ω)** | **Sample Resistance (Ω)** | **Noise** | **Sample Noise to Coil Noise Ratio** |
| **Room-temperature Copper** | 218.44 | 158.45 | 1.378605238 | 44.25 | 63.61 | 5.49 | 2.08 | 47.64 | **0.378871** |
| **Low-temperature Copper** | 486.25 | 266.69 | 1.823277963 | 46 | 61.24 | 2.37 | 1.95 | 27.72 | **0.822785** |
| **Superconductor** | 5809.9 | 652.29 | 8.9069279 | 48 | 58.64 | 0.19 | 1.5 | 21.24 | **7.894737** |
| | | | | **Human Brain** | | | | | |
| **Material** | **Unloaded Q-factor** | **Loaded Q-factor** | **Unloaded to Loaded Q-factor Ratio** | **Capacitance (pF)** | **Inductance (μH)** | **Coil Resistance (Ω)** | **Sample Resistance (Ω)** | **Noise** | **Sample Noise to Coil Noise Ratio** |
| **Room-temperature Copper** | 218.44 | 138.88 | 1.572868664 | 42.75 | 66.11 | 5.69 | 3.26 | 51.82 | **0.572935** |
| **Low-temperature Copper** | 486.25 | 213.99 | 2.272302444 | 44 | 63.99 | 2.48 | 3.16 | 33.72 | **1.274194** |
| **Superconductor** | 5809.9 | 347.87 | 16.70135395 | 45.3 | 61.84 | 0.2 | 3.16 | 30.77 | **15.8** |

## 3.3 $B_1^+$ Evaluation

The B1+ field distributions were evaluated in the YZ-sagittal plane of the elbow, knee, and brain samples for the three conductive materials. Five 10 by 10 mm² regions of interest (ROIs) were used to evaluate the mean B1+ values, and these mean values from the ROIs were then used to generate a mean B1+ value for the YZ-sagittal slice.

For the human brain sample, the mean B1+ efficiency with room-temperature copper was 13.57 $\mu T/\sqrt{W}$. With low-temperature copper, the mean B1+ efficiency was 16.932 $\mu T/\sqrt{W}$, which corresponds to a 24.82% gain. The superconducting material resulted in a mean B1+ efficiency of 21.814 $\mu T/\sqrt{W}$, showing a 60.74% gain compared to room-temperature copper (see figure 4a). For the human elbow, the mean B1+ efficiency with room-temperature copper was 15.962 $\mu T/\sqrt{W}$. With low-temperature copper, the mean B1+ efficiency increased to 21.846 $\mu T/\sqrt{W}$, representing a 36.82% gain. The superconducting material showed a mean B1+ efficiency of 42.174 $\mu T/\sqrt{W}$, which is a 164.14% gain compared to room-temperature copper (see figure 4b).

In the human knee sample, the mean B1+ efficiency with room-temperature copper was 14.09 $\mu T/\sqrt{W}$. For low-temperature copper, the mean B1+ efficiency was 17.952 $\mu T/\sqrt{W}$, reflecting a 27.42% gain. The superconducting material achieved a mean B1+ efficiency of 28.93 $\mu T/\sqrt{W}$, indicating a 105.32% gain over room-temperature copper (see figure 4c). These results demonstrate that the B1+ efficiency improves significantly with higher conductivity materials, as summarized in Table 2.

*Table 2*. Mean B1 efficiency and gain relative to room temperature copper for various conductivity parameters across different human models.

| Human Elbow | | |
|---|---|---|
| **Material** | **Mean $B_1^+$ Efficiency** | **$B_1^+$ Efficiency gain with respect to Room-temperature Copper** |
| **Room-temperature Copper** | 15.962 | 0% |
| **Low-temperature Copper** | 21.846 | 36.8% |
| **Superconductor** | 42.174 | 164.1% |

| Human Knee | | |
|---|---|---|
| **Material** | **Mean $B_1^+$ Efficiency** | **$B_1^+$ Efficiency gain with respect to Room-temperature Copper** |
| **Room-temperature Copper** | 14.09 | 0% |
| **Low-temperature Copper** | 17.952 | 27.4% |
| **Superconductor** | 28.93 | 105.2% |

| Human Brain | | |
|---|---|---|
| **Material** | **Mean $B_1^+$ Efficiency** | **$B_1^+$ Efficiency gain with respect to Room-temperature Copper** |
| **Room-temperature Copper** | 13.57 | 0% |
| **Low-temperature Copper** | 16.932 | 24.8% |
| **Superconductor** | 21.814 | 60.75% |

**B1+ Efficiency**

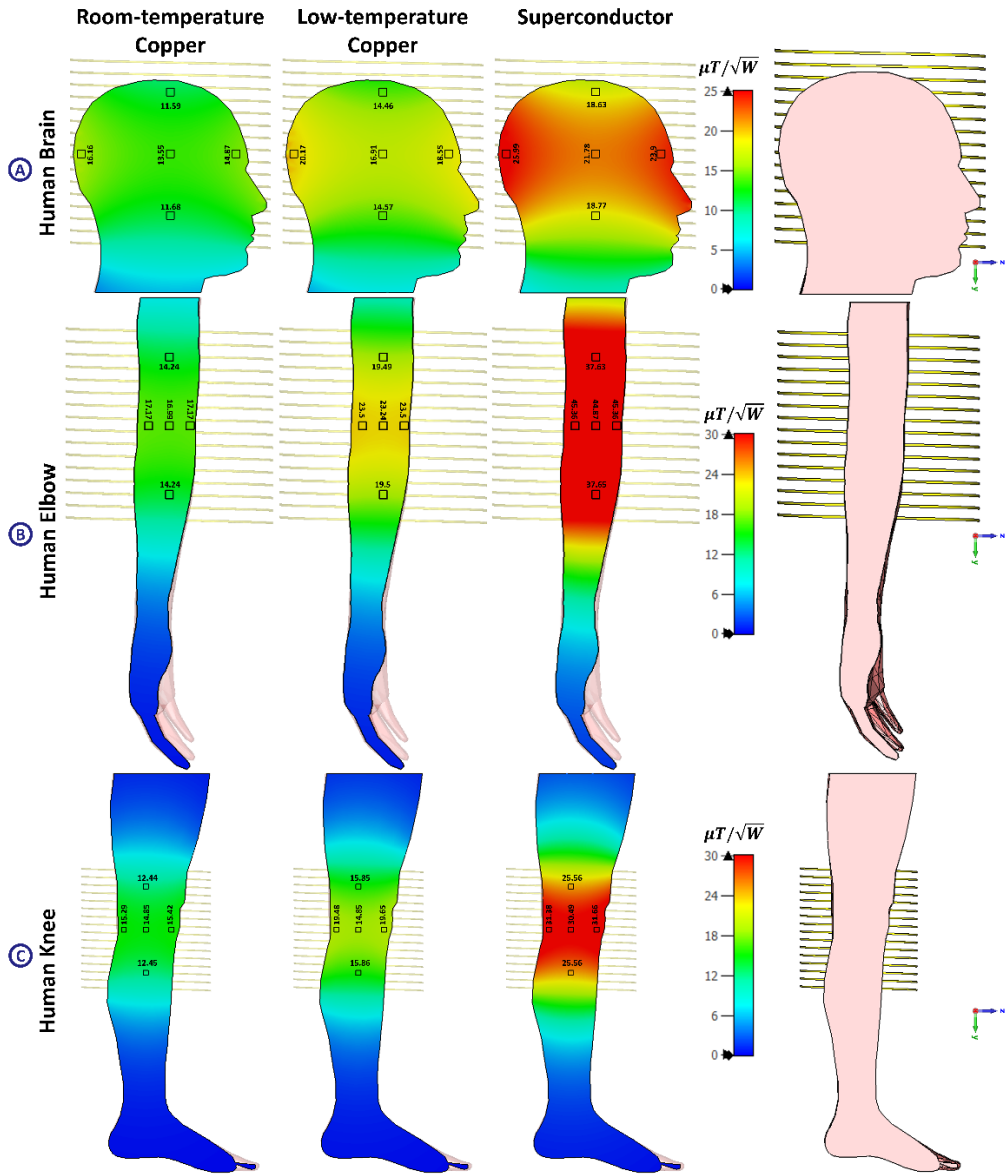

*Figure 4.* B1 efficiency in the central sagittal YZ-plane in $\mu T/\sqrt{W}$ for different conductivity parameters: (a) human brain, (b) human elbow, (c) human knee.

## 3.4 SNR Evaluation

The Signal-to-Noise Ratio (SNR) evaluation was performed using the B1+ field maps as the signal input and the noise values calculated from the previous section. The SNR is directly proportional to the ratio of the B1+ field to the noise. Therefore, the SNR maps were generated to assess the coil's performance with different conductive materials. The mean SNR values for the human elbow were 0.36881 for room-temperature copper, 0.84280 for low-temperature copper, and 2.8366 for the superconducting material. The SNR gain with respect to room-temperature copper was 128.57% for low-temperature copper and 669.21% for the superconducting material (see figure 5).

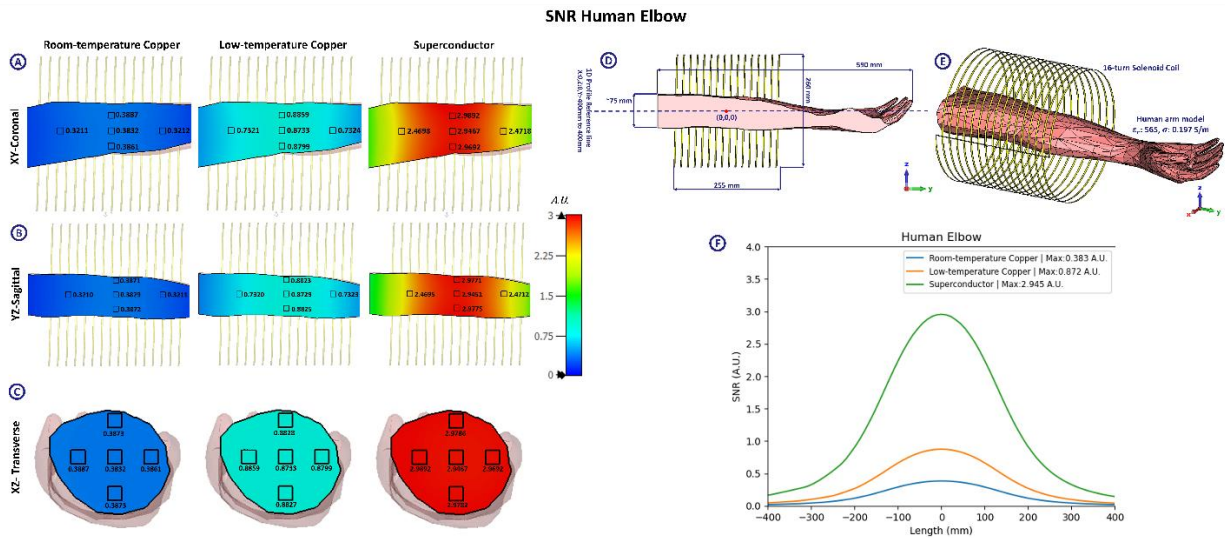

*Figure 5.* (a) SNR in A.U. for the human elbow in the central XY-coronal plane. (b) SNR in A.U. for the human elbow in the central YZ-sagittal plane. (c) SNR in A.U. for the human elbow in the central XZ-transverse plane. (d) Side view of the setup in the central YZ-sagittal plane, showing human arm model dimensions, coil dimensions, and the 1D profile reference line used to plot the SNR 1D profile. (e) 3D

*perspective view of the solenoid coil with the human arm model and labeled material properties. (f) 1D profile of the SNR in the human elbow for different conductivity parameters.*

For the human knee, the mean SNR values were 0.31082 for room-temperature copper, 0.6976 for low-temperature copper, and 1.3971 for the superconducting material. The SNR gain with respect to room-temperature copper was 124.41% for low-temperature copper and 349.45% for the superconducting material (see figure 6).

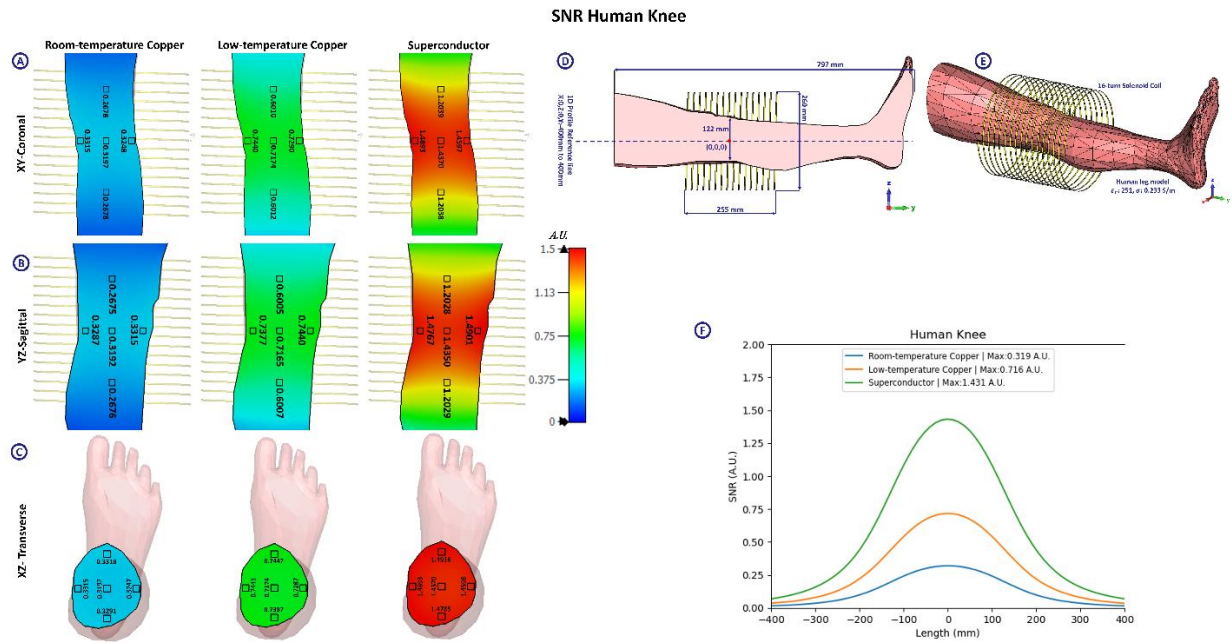

***Figure 6.*** *(a) SNR in A.U. for the human knee in the central XY-coronal plane. (b) SNR in A.U. for the human knee in the central YZ-sagittal plane. (c) SNR in A.U. for the human knee in the central XZ-transverse plane. (d) Side view of the setup in the central YZ-sagittal plane, showing human leg model dimensions, coil dimensions, and the 1D profile reference line used to plot the SNR 1D profile. (e) 3D perspective view of the solenoid coil with the human leg model and labeled material properties. (f) 1D profile of the SNR in the human knee for different conductivity parameters.*

In the human brain, the mean SNR values were 0.28278 for room-temperature copper, 0.5327 for low-temperature copper, and 0.72758 for the superconducting material. The SNR gain with respect to room-temperature copper was 88.43% for low-temperature copper and 157.45% for the superconducting material (see figure 7). These SNR evaluations highlight the impact of different conductivity materials on the coil's performance, demonstrating that higher conductivity materials result in significantly improved SNR, which is crucial for high-quality MR imaging. The details of the mean SNR and gain relative to room-temperature copper for various conductivity parameters across different human models are summarized in Table 3.

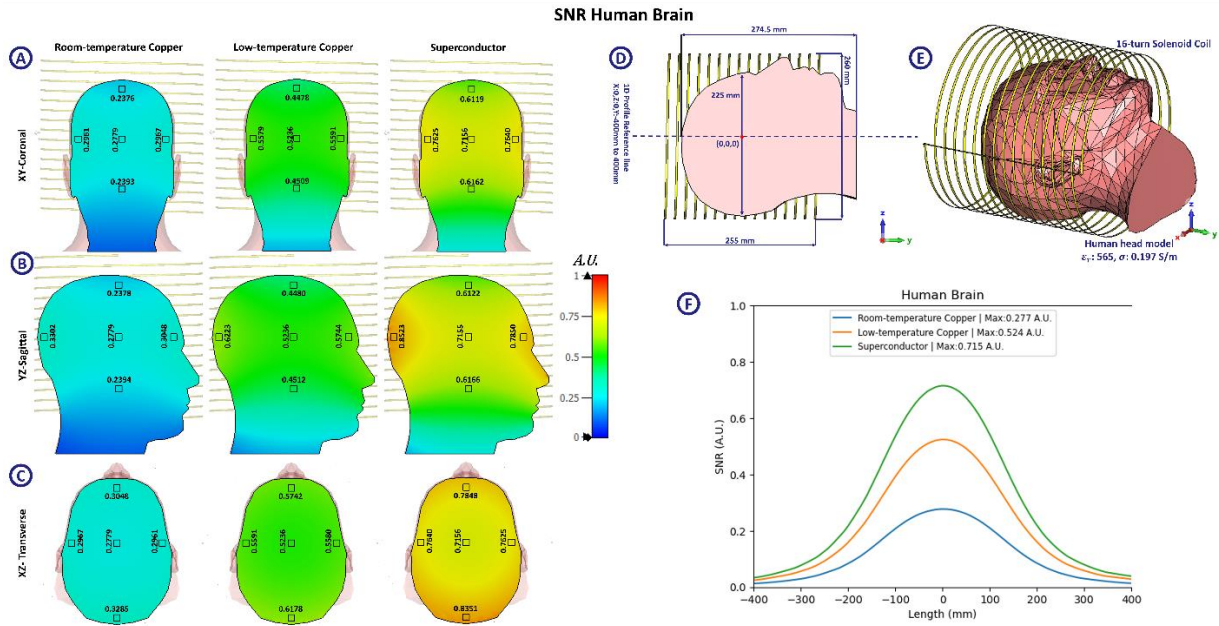

*Figure 7.* *(a) SNR in A.U. for the human brain in the central XY-coronal plane. (b) SNR in A.U. for the human brain in the central YZ-sagittal plane. (c) SNR in A.U. for the human brain in the central XZ-transverse plane. (d) Side view of the setup in the central YZ-sagittal plane, showing human head model dimensions, coil dimensions, and the 1D profile reference line used to plot the SNR 1D profile. (e) 3D perspective view of the solenoid coil with the human head model and labeled material properties. (f) 1D profile of the SNR in the human brain for different conductivity parameters.*

*Table 3.* *Mean SNR and gain relative to room temperature copper for various conductivity parameters across different human models.*

| Human Elbow | | |
|---|---|---|
| **Material** | **Mean SNR** | **SNR Efficiency gain with respect to Room-temperature Copper** |
| **Room-temperature Copper** | 0.36881 | 0% |
| **Low-temperature Copper** | 0.84280 | +128.5% |
| **Superconductor** | 2.8366 | +669.1% |
| Human Knee | | |
| **Material** | **Mean SNR** | **SNR Efficiency gain with respect to Room-temperature Copper** |
| **Room-temperature Copper** | 0.31082 | 0% |
| **Low-temperature Copper** | 0.6976 | +124.4% |
| **Superconductor** | 1.3971 | +349.4% |
| Human Brain | | |
| **Material** | **Mean SNR** | **SNR Efficiency gain with respect to Room-temperature Copper** |
| **Room-temperature Copper** | 0.28278 | 0% |
| **Low-temperature Copper** | 0.5327 | +88.3% |
| **Superconductor** | 0.72758 | +157.2% |

# Discussions & Conclusion

This study evaluated the performance of multi-turn solenoid coils designed for ultra-low field MR imaging, focusing on Q-factor, B1+ field efficiency, and SNR using different conductor materials (superconducting material, low-temperature copper, and room-temperature copper) across various human samples (elbow, knee, and brain). The results revealed that the superconducting material significantly outperformed both room-temperature and low-temperature copper at 70mT. The high Q-factor of the superconducting material indicates much lower energy loss, which is advantageous for MR imaging. Additionally, the B1+ field efficiency increased substantially with the superconducting material, highlighting its superior performance.

The SNR evaluations indicated that higher conductivity materials significantly enhance the SNR of the coil. The substantial improvements in SNR are crucial for achieving high-quality MR images, as they directly correlate with image clarity and detail. The findings have important implications for the design and optimization of MR imaging coils. The use of superconducting materials offers significant performance advantages despite the challenges associated with cooling and maintaining superconducting conditions. The improvements in Q-factor, B1+ efficiency, and SNR suggest that superconducting and low temperature coils could substantially enhance the quality of MR images at ultra-low fields, particularly in applications requiring high sensitivity and resolution.

In conclusion, this study demonstrates that at ultra-low fields, the MR noise is still dominated by the noise of RF coils, and that the performance of a multi-turn solenoid coil for MR imaging at ultra-low fields can be significantly enhanced by using superconducting materials and low temperature copper. Future research should explore practical implementation strategies for superconducting coils in clinical settings and the potential benefits in specific MR imaging scenarios. This study provides a strong foundation for further investigation into using high-conductivity materials in ultra-low field MR RF coil design, aiming to achieve superior image quality and diagnostic capabilities.